\documentclass[12pt]{article}
\usepackage{hyperref}
\usepackage[english]{babel}
\usepackage{amsmath}
\usepackage{latexsym}
\usepackage{amssymb}
\usepackage[all]{xy}
\usepackage[dvips]{graphicx}
\usepackage{epsfig}
\usepackage{psfrag}

\numberwithin{equation}{section} \numberwithin{table}{section}
\numberwithin{figure}{section}

\begin{document}
%\setlength{\parskip}{10pt}
%\oddsidemargin=0.5cm \evensidemargin=0.5cm \hoffset 0.6cm

%\voffset 0.0cm \hoffset 1.0cm \topmargin 0.0in \evensidemargin 0.0in
%\oddsidemargin 0.0in \textheight 8.6in \textwidth 7.25in
%\parskip 10 pt \pagestyle{plain}

%%%%%%%%%%%%%%%%%%%%%%%%%%%%%%%%%%%%%%%%%%%

%%%%%%%%%%%%%%%%%%%%%%%%%%%%%%%%%%%%%%%%%%%%%%%%%%%%%%%%%%%%%%%%%%%%%%%%
% title page
%%%%%%%%%%%%%%%%%%%%%%%%%%%%%%%%%%%%%%%%%%%%%%%%%%%%%%%%%%%%%%%%%%%
\begin{titlepage}
   %\begin{flushright}
%  {\small CQUeST-2011-xxxx}
%  \end{flushright}

   \begin{center}

     \vspace{20mm}

     {\LARGE \bf Holographic charge transport}\\
     \vspace{3mm}
     {\LARGE \bf in Lifshitz black hole backgrounds}

     \vspace{10mm}

    %Rong-Gen Cai$^{\ast}$ and Da-Wei Pang$^{\sharp\ast}$
    Jos\'{e} P. S. Lemos and Da-Wei Pang

     \vspace{5mm}
     %{\small \sl $\ast$ Key Laboratory of Frontiers in Theoretical
%Physics,\\
%    Institute of Theoretical Physics, Chinese Academy of Sciences, \\
%    P.O. Box 2735, Beijing 100190, China\\}

      {\small \sl  Centro Multidisciplinar de Astrof\'{\i}sica-CENTRA,
      Departamento de F\'{\i}sica,}\\
     {\small \sl Instituto Superior T\'{e}cnico-IST,
      Universidade T\'{e}cnica de Lisboa}\\
     {\small \sl Av. Rovisco Pais 1, 1049-001 Lisboa, Portugal}\\
     %{\small \tt cairg@itp.ac.cn, dwpang@gmail.com}
     {\small \tt joselemos@ist.utl.pt, dawei.pang@ist.utl.pt}
    % {\small \tt dwpang@gmail.com}
     \vspace{10mm}

   \end{center}

\begin{abstract}
\baselineskip=18pt
We study charge transport properties in a domain-wall geometry, whose
near horizon IR geometry is a Lifshitz black hole and whose UV
geometry is AdS. The action for the gauge field contains the standard
Maxwell term plus the Weyl tensor coupled to Maxwell field strengths.
In four dimensions we calculate the conductivity via both the membrane
paradigm and Kubo's formula. Precise agreements between both methods
are obtained. Moreover, we perform an analysis of the
four-dimensional electro-magnetic duality in our domain-wall
background and find that the relation between the longitudinal and
transverse components of the current-current correlation functions and
those of the `dual' counterparts holds, irrespective of the near
horizon IR geometry. Conductivity at extremality is also
investigated. Generalizations to higher dimensions are performed.

\end{abstract}
\setcounter{page}{0}
\end{titlepage}

\pagestyle{plain} \baselineskip=19pt

\tableofcontents

%%%%%%%%%%%%%%%%%%%%%%%%%%%%%%%%%%%%%%%%%%%%%%
\section{Introduction}
%%%%%%%%%%%%%%%%%%%%%%%%%%%%%%%%%%%%%%%%%%%%%%

The AdS/CFT correspondence~\cite{Maldacena:1997re, Aharony:1999ti},
a duality between gravity (AdS) and gauge field theory (CFT),
also called the gauge/gravity duality, sets up connections between
gravity theory in a certain bulk spacetime and field theory on the
boundary of that spacetime. It has been widely recognized that the
gauge/gravity duality provides powerful tools for studying dynamics of
strongly coupled field theories and physics in the real
world. Recently, investigations on applications of the AdS/CFT
correspondence to condensed matter physics (AdS/CMT for short) have,
due to its great interest,
increased enormously~\cite{Hartnoll:2009sz}. For instance, gravity
backgrounds which possess non-relativistic symmetries were constructed
in~\cite{Son:2008ye, Balasubramanian:2008dm, Kachru:2008yh}.

One crucial quantity characterizing charge transport properties of
condensed matter systems is the conductivity, which can be evaluated
via the current-current correlation function of the bulk $U(1)$ gauge
field in the dual gravity side. It was found in~\cite{Herzog:2007ij}
that the conductivity in the three-dimensional field theory side at
zero momentum was a constant with no frequency dependence. The
authors of~\cite{Herzog:2007ij} attributed this remarkable result to
the electro-magnetic duality of the four-dimensional bulk
Einstein-Maxwell theory. Recently in order to acquire a better
understanding of this self-duality, Myers, Sachdev and
Singh~\cite{Myers:2010pk} considered a particular form of new higher
derivative corrections which involves couplings between the gauge
field to the spacetime curvature. The higher order corrections to the
conductivity were obtained and they found that although the
electro-magnetic self-duality was lost in the presence of higher order
corrections, a simple relation between the transverse and longitudinal
components of the current-current retarded correlation function and
those of the `dual' counterparts still held.

Since many condensed matter systems possess non-relativistic
symmetries, it is desirable to study the conductivity in such
non-relativistic backgrounds and to see if the `duality' relation for
the current-current correlation functions still holds. In this paper
we consider charge transport properties at Lifshitz fixed points. The
background is a domain wall geometry, where the metric becomes
a Lifshitz black hole in the IR and an asymptotically AdS spacetime
in the UV. The action for the bulk $U(1)$ gauge field contains the
ordinary Maxwell term, as well as coupling between the Weyl tensor and
the field strengths. First we work in four dimensions and calculate
the conductivity via the membrane paradigm, which reduces to the one
obtained in~\cite{Myers:2010pk} when the dynamical exponent
$z=1$. Next we evaluate the conductivity from Kubo's formula, which
precisely matches the result obtained via the membrane paradigm.
Moreover, we find that the relation between the transverse and
longitudinal components of the current-current retarded correlation
functions and those of the `dual' counterparts still holds,
irrespective of the IR geometry. We also comment on the
conductivity at extremality. Generalizations to higher-dimensional
spacetimes are also obtained.

The rest of the paper is organized as follows: we give a brief review
on relevant backgrounds in section~\ref{sec2}.  Then we focus on
charge transport properties in four dimensions in
section~\ref{sec3}. Firstly we calculate the conductivity using the
membrane paradigm in section~\ref{sec3.1}, where we find that although
the Weyl corrections do not contribute in the $z=2$ case, it is indeed
a coincidence which can be seen by considering more general
actions. Next in section~\ref{sec3.2} we reconsider the conductivity
by evaluating the retarded current-current correlation functions and
find precise agreement with the result obtained in
section~\ref{sec3.1}. A simple relation between the transverse and
longitudinal components of the current-current retarded correlation
functions and those of the `dual' counterparts is derived in
section~\ref{sec3.3}, which agrees with that obtained
in~\cite{Myers:2010pk}.
In section~\ref{sec3.4} conductivity at extremality is investigated.
Higher-dimensional generalizations are
evaluated in section~\ref{sec4} and discussions on other
related issues are given in section~\ref{sec5}.

%%%%%%%%%%%%%%%%%%%%%%%%%%%%%%%%%%%%%%%%%%%%%%%%%%%%%%%%%
\section{Preliminaries}
\label{sec2}
%%%%%%%%%%%%%%%%%%%%%%%%%%%%%%%%%%%%%%%%%%%%%%%%%%%%%%%%%%
In this section we review some relevant backgrounds before
proceeding. First of all, the starting point in~\cite{Myers:2010pk}
was the four-dimensional planar Schwarzschild-${\rm AdS}_{4}$ black
hole \cite{Lemos:1994fn},
\begin{equation}
\label{metricplana}
ds^{2}=\frac{r^{2}}{L^{2}}(-f(r)dt^{2}+dx^{2}+dy^{2})+
\frac{L^{2}dr^{2}}{r^{2}f(r)}\,,
\end{equation}
where $f(r)=1-r^{3}_{0}/r^{3}$. On the other hand,
after integrating by parts and imposing
the identities $\nabla_{[a}F_{bc]}=R_{[abc]d}=0$,
the most general four-derivative action
contains the following terms,
\begin{eqnarray}
\label{equ1}
I_{4}&=&\int d^{4}x\sqrt{-g}[\alpha_{1}R^{2}+\alpha_{2}
R_{ab}R^{ab}+\alpha_{3}(F^{2})^{2}+\alpha_{4}F^{4}
+\alpha_{5}\nabla^{a}F_{ab}\nabla^{c}{F_{c}}^{b}\nonumber\\
& &~~~~~~~~~~~~~+\alpha_{6}R_{abcd}F^{ab}F^{cd}+
\alpha_{7}R^{ab}F_{ac}{F_{b}}^{c}+\alpha_{8}RF^{2}],
\end{eqnarray}
where $F^{2}=F_{ab}F^{ab},
F^{4}={F^{a}}_{b}{F^{b}}_{c}{F^{c}}_{d}{F^{d}}_{a}$.  If we focus on
the conductivity, which means that only the current-current two-point
functions are relevant, we can just consider the effects of the
$\alpha_{6}, \alpha_{7}$ and $\alpha_{8}$ terms. Furthermore, after
taking a particular linear combination of these three terms, the
effective action for bulk Maxwell field turns out to be
\begin{equation}
\label{equ2}
I_{\rm vec}=\frac{1}{g^{2}_{4}}\int
d^{4}x\sqrt{-g}[-\frac{1}{4}F_{ab}F^{ab}+\gamma
L^{2}C_{abcd}F^{ab}F^{cd}],
\end{equation}
where $C_{abcd}$ denotes the Weyl tensor. One advantage of taking this
particular combination is that the asymptotic geometry will not be
modified, as the Weyl tensor vanishes in pure AdS space. Then the DC
conductivity in the presence of higher order corrections is given by
\begin{equation}
\sigma_{\rm DC}=\frac{1}{g^{2}_{4}}(1+4\gamma).
\end{equation}

We shall consider the following domain-wall geometry
\begin{equation}
\label{dmwall}
ds^{2}=-g(r)e^{-\chi(r)}dt^{2}+
\frac{dr^{2}}{g(r)}+\frac{r^{2}}{R_{0}^2}(dx^{2}+dy^{2}),
\end{equation}
where $R_{0}$ denotes certain length scale.
The IR region is described by a Lifshitz black hole
(see, e.g., \cite{Taylor:2008tg, Lif}),
\begin{equation}
\label{4dIR}
ds^{2}_{\rm
IR}=-\frac{r^{2z}}{L^{2}}f(r)dt^{2}+
\frac{L^{2}dr^{2}}{r^{2}f(r)}+
\frac{r^{2}}{L^{2}}(dx^{2}+dy^{2}),
~~~f(r)=1-\frac{r_{0}^{z+2}}{r^{z+2}},
\end{equation}
where $z$ is the dynamical exponent. The above background possesses
the following Lifshitz scaling symmetry at extremality when $f(r)=1$,
\begin{equation}
t\rightarrow\lambda^{z}t,~~~r\rightarrow
\frac{r}{\lambda},~~~\vec{x}\rightarrow\lambda\vec{x}.
\end{equation}
Generically, such solutions
are always accompanied by various matter fields and the form of
$f(r)$ is determined by the matter fields.
However, here we just write down the metric as above so that it
becomes Schwarzschild-${\rm AdS}_{4}$ when $z=1$.
Combining~(\ref{dmwall}) and~(\ref{4dIR}), we can find that
\begin{equation}
e^{-\chi(r)}=r^{2z-2},~~~g(r)=\frac{r^{2}f(r)}{L^{2}},~~~R_{0}=L.
\end{equation}
The UV geometry is chosen to be AdS so that it will not be modified by
the higher order corrections~(\ref{equ2}) and we can still perform
calculations in the context of AdS/CFT. Such a domain-wall geometry
holographically describes a RG flow towards a nontrivial IR Lifshitz
fixed point.

In this paper the action for the Maxwell field is still given
by~(\ref{equ2}) and the equation of motion
reads
\begin{equation}
\nabla_{a}[F^{ab}-4\gamma L^{2}C^{abcd}F_{cd}]=0.
\end{equation}
We also list the non-vanishing components of the Weyl
tensor for later convenience
\begin{eqnarray}
\label{4dweyl}
& &C_{trtr}=\frac{e^{-\chi(r)}}{12r^{2}}F(r),~~~
C_{titj}=-\frac{e^{-\chi(r)}}{24R_{0}^2}g(r)F(r)\delta_{ij},\nonumber\\
& &C_{rirj}=\frac{1}{24R_{0}^{2}}\frac{F(r)}{g(r)}\delta_{ij},~~~
C_{ijkl}=-\frac{r^{2}}{12R_{0}^{4}}F(r)\delta_{ik}\delta_{jl},
\end{eqnarray}
where $i,j,k,l=x,y$ and
\begin{eqnarray}
\label{fr}
F(r)&=&r[-g^{\prime}(r)(4+
3\chi^{\prime}(r))+2rg^{\prime\prime}(r)]\nonumber\\
&
&+g(r)(4+2r\chi^{\prime}(r)+
r^{2}\chi^{\prime2}(r)-2r^{2}\chi^{\prime\prime}(r)).
\end{eqnarray}
%%%%%%%%%%%%%%%%%%%%%%%%%%%%%%%%%%%%%%%%%%
\section{Charge transport in four dimensions}
\label{sec3}
%%%%%%%%%%%%%%%%%%%%%%%%%%%%%%%%%%%%%%%%%%%
We study charge transport properties in a four-dimensional domain-wall
background, which is the most interesting case. In
section~\ref{sec3.1} we calculate the conductivity using the membrane
paradigm and verify the result via Kubo's formula in
section~\ref{sec3.2}. A simple relation between the longitudinal and
transverse parts of the current-current correlation functions and
those of the `dual' counterparts is derived in section~\ref{sec3.3}.
In addition, we briefly discuss the conductivity at zero temperature
in section~\ref{sec3.4}.
%%%%%%%%%%%%%%%%%%%%%%%%%%%%%%%%%%%%%%%%%%%%%%
\subsection{DC conductivity from the membrane paradigm}
\label{sec3.1}
%%%%%%%%%%%%%%%%%%%%%%%%%%%%%%%%%%%%%%%%%%%%%%
In this subsection, we calculate the DC conductivity via the membrane
paradigm, following ~\cite{Brigante:2007nu, Ritz:2008kh}. Such a
prescription can be seen as a generalization of the analysis
in~\cite{Kovtun:2003wp, Iqbal:2008by} to incorporate the following
general action
\begin{equation}
\label{xabcd}
I=\int d^{4}x\sqrt{-g}(-\frac{1}{8g^{2}_{4}}F_{ab}X^{abcd}F_{cd}),
\end{equation}
where the tensor $X^{abcd}$ possesses the following symmetries
$X^{abcd}=X^{[ab][cd]}=X^{cdab}$.
For our particular example,
\begin{equation}
{X_{ab}}^{cd}={I_{ab}}^{cd}-8\gamma L^{2}{C_{ab}}^{cd},
\end{equation}
where
\begin{equation}
{I_{ab}}^{cd}={\delta_{a}}^{c}{\delta_{b}}^{d}-
{\delta_{a}}^{d}{\delta_{b}}^{c},
\end{equation}
so that the above action reduces to the conventional Maxwell action
when $\gamma=0$.

Extensions to the general action~(\ref{xabcd}) are straightforward. We
still define the stretched horizon at $r=r_{H}$, where $r_{H}>r_{0}$
and $r_{H}-r_{0}\ll r_{0}$.  The corresponding conserved current is
given by
\begin{equation}
j^{a}=\frac{1}{4}n_{b}X^{abcd}F_{cd}|_{r=r_{0}},
\end{equation}
where $n_{a}$ is an ourward-pointing radial unit vector.
According to Ohm's law at the stretched horizon, the DC conductivity
reads
\begin{equation}
\sigma=\frac{1}{g^{2}_{4}}\sqrt{-g}
\sqrt{-X^{txtx}X^{rxrx}}|_{r=r_{0}}.
\end{equation}
Plugging~(\ref{4dweyl}) and~(\ref{fr}) into
the above expression, we can arrive at
\begin{equation}
\label{dc4}
\sigma=\frac{1}{g^{2}_{4}}[1-\frac{4}{3}\gamma(z^{2}-4)].
\end{equation}
When $z=1$, the conductivity turns out to be
\begin{equation}
\label{dcs}
\sigma=\frac{1}{g^{2}_{4}}[1+4\gamma]\,,
\end{equation}
which agrees with that obtained~\cite{Myers:2010pk}.

It can be easily seen that when $z=2$, $\sigma=1/g^{2}_{4}$, which
means that the conductivity is not corrected by the higher order
terms. One may wonder if this fact implies some underlying physics or
just a coincidence. To answer this question, we can consider a more
general form of corrections
\begin{equation}
\tilde{I}_{\rm vec}=\frac{1}{g^{2}_{4}}\int
d^{4}x\sqrt{-g}[-\frac{1}{4}F_{ab}F^{ab}+\gamma
L^{2}(c_{1}R_{abcd}F^{ab}F^{cd}+c_{2}R_{ab}F^{ac}{F^{b}}_{c}+
c_{3}RF^{ab}F_{ab})],
\end{equation}
where $c_{i}, i=1,2,3$ are constants. Now the tensor $X^{abcd}$
becomes
\begin{eqnarray}
\tilde{X}^{abcd}&=&(g^{ac}g^{bd}-g^{ad}g^{bc})-8\gamma
L^{2}[c_{1}R^{abcd}\nonumber+\frac{c_{2}}{4}
(R^{ac}g^{bd}-R^{ad}g^{bc}+R^{bd}g^{ac}-R^{bc}g^{ad})\nonumber\\&
&+\frac{c_{3}}{2}R(g^{ac}g^{bd}-g^{ad}g^{bc})],
\end{eqnarray}
and the conductivity is given by
\begin{equation}
\tilde{\sigma}=\frac{1}{g^{2}_{4}}\sqrt{-g}
\sqrt{-\tilde{X}^{txtx}\tilde{X}^{rxrx}}|_{r=r_{0}}
=\frac{1}{g^{2}_{4}}[1+2\gamma(z+2)(2c_{1}+(c_{2}+4c_{3})(z+1))].
\end{equation}
It can be seen that $z=2$ also leads to nontrivial higher order
corrections for general $c_{i}$'s. In particular, when $c_{1}=1,
c_{2}=-2, c_{3}=1/3$, the tensor
$\tilde{X}^{abcd}=(g^{ac}g^{bd}-g^{ad}g^{bc})-8\gamma L^{2}C^{abcd}$,
and the conductivity is given by
\begin{equation}
\tilde{\sigma}=\frac{1}{g^{2}_{4}}[1-\frac{4}{3}\gamma(z^{2}-4)],
\end{equation}
which agrees with~(\ref{dc4}). Hence the `non-renormalization' of the
conductivity is just due to our particular choice of the higher order
corrections.

The membrane paradigm also determines the charge diffusion constant
\begin{equation}
D=-\sqrt{-g}\sqrt{-X^{txtx}X^{rxrx}}|_{r=r_{0}}\int^{\infty}_{r_{0}}
\frac{dr}{\sqrt{-g}X^{trtr}}.
\end{equation}
However, here we cannot evaluate the charge diffusion constant in a
similar way, as we are studying the domain-wall geometry and we only
explicitly know the IR and the UV geometries. It can be seen that the
$r\rightarrow\infty$ limit of ~(\ref{4dIR}) leads to Lifshitz metric
rather than AdS metric, which means that we cannot calculate the
charge diffusion constant by naively applying this formula.
%\begin{figure}
%\begin{center}
%\vspace{3cm} \hspace{-0.5cm}
%\includegraphics[angle=0,width=0.4\textwidth]{fig1.eps}
%\hspace{0.5cm}
%\includegraphics[angle=0,width=0.4\textwidth]{fig2.eps}
%%\vspace{-2.5cm} \\
%\caption{\small The specific heat (blue line) and electrical
%permittivity (purple line).
%Left: $\alpha=0.01, Q=0.1r_{0}$. Right: $\alpha=0.01, Q=0.5r_{0}$.}
%\end{center}
%\end{figure}
%%%%%%%%%%%%%%%%%%%%%%%%%%%%%%%%%%%%%%%%%%%%%%%
\subsection{DC Conductivity from Kubo's formula}
\label{sec3.2}
%%%%%%%%%%%%%%%%%%%%%%%%%%%%%%%%%%%%%%%%%%%%%%%%
In this subsection, we reconsider the DC conductivity by making use of
Kubo's formula, which can be seen as a check of consistency for the
result obtained via the membrane paradigm. According to Kubo's
formula, in the hydrodynamic limit the conductivity can be determined
in terms of the retarded current-current correlation function
\begin{equation}
\label{4eq1}
\sigma_{\rm DC}=-\lim_{\omega\rightarrow0}
\frac{1}{\omega}{\rm Im}G^{R}_{xx}(\omega,\vec{k}=0),
\end{equation}
where
\begin{equation}
G^{R}_{xx}(\omega,\vec{k}=0)=
-i\int dtd\vec{x}e^{i\omega t}\theta(t)
\langle[J_{x}(x),J_{x}(0)]\rangle.
\end{equation}
Here $J_{x}$ denotes the CFT current dual to
the bulk gauge field $A_{x}$. In this subsection we also
introduce a new radial coordinate $u$, in which
the domain wall metric can be written as
\begin{equation}
\label{uzb}
ds^{2}=-g(u)e^{-\chi(u)}dt^{2}+\frac{du^{2}}{g(u)}+
\frac{R_{0}^{2}}{u^{2}}(dx^{2}+dy^{2}).
\end{equation}
The IR geometry can be expressed as
\begin{equation}
\label{4diru}
ds^{2}_{\rm
IR}=-\frac{r_{0}^{2z}}{L^{2}u^{2z}}f(u)dt^{2}+\frac{L^{2}du^{2}}{u^{2}f(u)}
+\frac{r_{0}^{2}}{L^{2}u^{2}}(dx^{2}+dy^{2}),~~~f(u)=1-u^{z+2},
\end{equation}
where the horizon locates at $u=1$. Notice that $u=0$ does not
correspond to the asymptotic boundary. Comparing the above two metrics we
can find
\begin{equation}
e^{-\chi(u)}=\frac{r_{0}^{2z}}{u^{2z+2}},~~~
g(u)=\frac{u^{2}f(u)}{L^{2}},~~~R_{0}=\frac{r_{0}}{L}.
\end{equation}
The non-vanishing components of the Weyl tensor are given as follows
\begin{eqnarray}
\label{weylu}
& &C_{tutu}=\frac{e^{-\chi(u)}}{12u^{2}}F(u),~~~
C_{titj}=-
\frac{R_{0}^{2}e^{-\chi(u)}}{24u^{4}}g(u)F(u)\delta_{ij},\nonumber\\
& &C_{uiuj}=\frac{R_{0}^{2}}{24u^{4}}\frac{F(u)}{g(u)}\delta_{ij},~~~
C_{ijkl}=-\frac{R_{0}^{4}}{12u^{6}}F(u)\delta_{ik}\delta_{jl},
\end{eqnarray}
where $i,j,k,l=x,y$ and
\begin{eqnarray}
\label{fu}
F(u)&=&u[g^{\prime}(u)(4-3u\chi^{\prime}(u))+2ug^{\prime\prime}(u)]
\nonumber\\
&
&-g(u)(4+2u\chi^{\prime}(u)-u^{2}\chi^{\prime2}(u)+2u^{2}
\chi^{\prime\prime}(u)),
\end{eqnarray}

Consider gauge field fluctuations of the following form
\begin{equation}
\label{atux}
A_{a}(t,u,x)=\int\frac{d^{3}q}{(2\pi)^{3}}e^{-i\omega t+iqx}A_{a}(u,q),
\end{equation}
where we have chosen the three-momentum vector ${\bf
q}^{\mu}=(\omega,q,0)$ and the gauge
$A_{u}(u,q)=0$. From~(\ref{4eq1}), it can be seen that to calculate
the conductivity, it is sufficient to set $q=0$ in subsequent
calculations. The $y$-component of the generalized Maxwell
equation reads
\begin{equation}
\label{4day}
A^{\prime\prime}_{y}+\frac{M^{\prime}(u)}{M(u)}A^{\prime}_{y}+
\frac{e^{\chi(u)}}{g(u)^{2}}
\omega^{2}A_{y}=0,
\end{equation}
where
\begin{equation}
M(u)=(1-\frac{\gamma L^{2}}{3u^{2}}F(u))e^{-\chi(u)/2}g(u).
\end{equation}
On the other hand, the boundary action is given by
\begin{equation}
I_{y}=-\frac{1}{2g^{2}_{4}}\int d^{3}x
\sqrt{-g}g^{uu}g^{yy}(1-8\gamma L^{2}{C_{uy}}^{uy})
A_{y}\partial_{u}A_{y}|_{u\rightarrow u_{b}},
\end{equation}
where $u_{b}$ denotes the boundary, as the $u\rightarrow0$ limit
of~(\ref{4diru}) also leads to Lifshitz geometry rather than
AdS. Therefore the corresponding retarded Green's function is given
by~\cite{Son:2002sd}
\begin{equation}
G^{R}_{yy}=-\frac{1}{g^{2}_{4}}\sqrt{-g}g^{uu}g^{yy}(1-
8\gamma L^{2}{C_{uy}}^{uy})
\frac{A_{y}(u,-q)\partial_{u}A_{y}(u,q)}{A_{y}(u,-q)A_{y}(u,q)}
\big|_{u\rightarrow u_{b}}.
\end{equation}
It can be easily seen that
\begin{equation}
\sqrt{-g}g^{uu}g^{yy}(1-8\gamma L^{2}{C_{uy}}^{uy})=M(u),
\end{equation}

As argued previously, since we do not know the explicit domain-wall
metric, we cannot obtain concrete forms of the correlation
functions. However, as noted in~\cite{CaronHuot:2006te,
Atmaja:2008mt}, there exists a shortcut to calculate the
conductivity. First of all, for a general second order differential
equation
\begin{equation}
Y^{\prime\prime}(u)+A(u)Y^{\prime}(u)+B(u)Y(u)=0,
\end{equation}
there exists a conserved quantity
\begin{equation}
\label{qu}
Q(u)=e^{\int A}(\bar{Y}\partial_{u}Y-Y\partial_{u}\bar{Y}).
\end{equation}
For our case, the conserved quantity $Q(u)$ is given by
\begin{equation}
Q(u)=M(u)(\bar{A}_{y}\partial_{u}A_{y}-A_{y}\partial_{u}\bar{A}_{y}),
\end{equation}
thus the imaginary part of the correlation function turns out to be
\begin{equation}
{\rm Im}G^{R}_{yy}=-\frac{1}{2ig^{2}_{4}}
\frac{Q(u)}{|A_{y}(u)|^{2}}|_{u\rightarrow u_{b}},
\end{equation}
where again $u_{b}$ denotes the boundary.
The solution for $A_{y}$ can be written as
\begin{equation}
A_{y}(u)=(1-u)^{-\frac{i\omega}{4\pi T}}y(u),
\end{equation}
where the exponent $-i\omega/(4\pi T)$ is determined by
solving~(\ref{4day}) in the near horizon region and imposing the
incoming boundary condition. Since $Q(u)$ is a conserved quantity, we
can evaluate it at the horizon $u=1$. Therefore
\begin{equation}
{\rm Im}G^{R}_{yy}=-\frac{\omega}{g^{2}_{4}}
\left(1-\frac{4}{3}\gamma(z^{2}-4)\right)
\frac{|y(1)|^{2}}{|y(u_{b})|^{2}}.
\end{equation}
Moreover, in the low frequency limit, the solution
to~(\ref{4day}) is simply $y(u)=\rm const$.
Finally we arrive at
\begin{equation}
\sigma=-\frac{1}{\omega}{\rm Im}G^{R}_{yy}=
\frac{1}{g^{2}_{4}}[1-\frac{4}{3}\gamma(z^{2}-4)],
\end{equation}
which agrees with that obtained before.
%%%%%%%%%%%%%%%%%%%%%%%%%%%%%%%%%%%%%%%%%%%%
\subsection{EM duality in four dimensions}
\label{sec3.3}
%%%%%%%%%%%%%%%%%%%%%%%%%%%%%%%%%%%%%%%%%%%%%
In this subsection we discuss electro-magnetic duality in our
domain-wall background, which can be seen as extensions
of~\cite{Myers:2010pk} to more general cases. Generally speaking,
current conservation and spatial rotational invariance fix the
following general structure of the retarded Green's functions
\begin{equation}
G^{R}_{\mu\nu}({\bf q})=\sqrt{{\bf
q^{2}}}(P^{T}_{\mu\nu}K^{T}(\omega,q)+P^{L}_{\mu\nu}K^{L}(\omega,q)),
\end{equation}
where ${\bf q}^{\mu}=(\omega,q^{x},q^{y}),
q^{2}=[(q^{x})^{2}+(q^{y})^{2}]^{1/2}, {\bf q}^{2}=q^{2}-\omega^{2}$.
Here $P^{T}_{\mu\nu}$ and $P^{L}_{\mu\nu}$ are orthogonal projection
operators given by
\begin{equation}
P^{T}_{tt}=P^{T}_{ti}=P^{T}_{it}=0,~~~P^{T}_{ij}=
\delta_{ij}-\frac{q_{i}q_{j}}{q^{2}},~~~
P^{L}_{\mu\nu}=\left(\eta_{\mu\nu}-
\frac{q_{\mu}q_{\nu}}{|{\bf q}|^{2}}\right)-P^{T}_{\mu\nu},
\end{equation}
where $i,j$ are spatial indices and $\mu,\nu$ denote the whole
spacetime indices. Let us choose ${\bf q}^{\mu}=(\omega,q,0)$ for
simplicity, then we have
\begin{equation}
G^{R}_{yy}(\omega,q)=\sqrt{q^{2}-\omega^{2}}K^{T}(\omega,q),~~~
G^{R}_{tt}(\omega,q)=-\frac{q^{2}}{\sqrt{q^{2}-\omega^{2}}}K^{L}(\omega,q)
\end{equation}
It was observed in~\cite{Herzog:2007ij} that at the leading order
level, i.e.  in the standard four-dimensional Maxwell theory, $K^{T}$
and $K^{L}$ satisfied the following simple relation
$$K^{T}(\omega,q)K^{L}(\omega,q)={\rm const},$$ which signifies
self-duality of the theory. As a result, the conductivity was a fixed
constant.

Following~\cite{Myers:2010pk}, we introduce a Lagrangian
multiplier $B_{a}$
\begin{equation}
I=\int d^{4}x\sqrt{-g}\left(-\frac{1}{8g^{2}_{4}}
F_{ab}X^{abcd}F_{cd}+
\frac{1}{2}\varepsilon^{abcd}B_{a}\partial_{b}F_{cd}\right),
\end{equation}
where $\varepsilon_{abcd}$ is the totally antisymmetric tensor with
$\epsilon_{0123}=\sqrt{-g}$.  After integrating by parts in the second
term and some other manipulations, the action can be written as
\begin{equation}
I=\int d^{4}x\sqrt{-g}\left(-\frac{1}{8\hat{g}^{2}_{4}}
\hat{X}^{abcd}G_{ab}G_{cd}\right),
\end{equation}
where $G_{ab}\equiv\partial_{a}B_{b}-\partial_{b}B_{a}$
denotes the new field
strength, $\hat{g}^{2}_{4}\equiv1/g^{2}_{4}$ and
\begin{equation}
{\hat{X}_{ab}}^{cd}=
-\frac{1}{4}{\varepsilon_{ab}}^{ef}{(X^{-1})_{ef}}^{gh}
{\varepsilon_{gh}}^{cd}.
\end{equation}
Here and in the following the hatted quantities denote those in the
`dual' theory. The field strengths $F_{ab}$ and $G_{ab}$ are related
by
\begin{equation}
\label{fg}
F_{ab}=\frac{g^{2}}{4}{(X^{-1})_{ab}}^{cd}{\varepsilon_{cd}}^{ef}G_{ef},
\end{equation}
In standard Maxwell theory, the two actions and the corresponding
equations of motion for $A_{a}$ and $B_{a}$ are identical, which means
that the Maxwell theory is self-dual. Moreover, the duality relation
between $F_{ab}$ and $G_{ab}$ is the usual Hodge dual.

In general $\hat{X}\neq X$, which means that self-duality is lost. The
corresponding equations of motion are given by
\begin{equation}
\nabla_{a}(X^{abcd}F_{cd})=0,~~~\nabla_{a}(\hat{X}^{abcd}G_{cd})=0.
\end{equation}
It can be seen that in the small $\gamma$ limit
\begin{equation}
{(X^{-1})_{ab}}^{cd}={I_{ab}}^{cd}+8\gamma
L^{2}{C_{ab}}^{cd}+O(\gamma^{2}).
\end{equation}
Furthermore, using traceless properties of the Weyl tensor, we obtain
\begin{equation}
{\hat{X}_{ab}}^{cd}={(X^{-1})_{ab}}^{cd}+O(\gamma^{2}),
\end{equation}
Introducing index pairs $A,B\in\{tx,ty,tu,xy,xu,yu\}$ we write
\begin{equation}
{X_{A}}^{B}={\rm diag}(X_{1},X_{2},X_{3},X_{4},X_{5},X_{6}),
\end{equation}
and
\begin{equation}
{\hat{X}_{A}}^{B}={\rm diag}\left(\frac{1}{X_{6}},
\frac{1}{X_{5}},\frac{1}{X_{4}},\frac{1}{X_{3}},
\frac{1}{X_{2}},\frac{1}{X_{1}}\right).
\end{equation}
The relation between $F_{ab}$ and $G_{ab}$~(\ref{fg}) becomes
\begin{equation}
F_{A}=g^{2}_{4}{(X^{-1})_{A}}^{B}{\varepsilon_{B}}^{C}G_{C}\,.
\end{equation}

Here our background is shown in~(\ref{uzb}) and the non-vanishing
components of the Weyl tensor are given in~(\ref{weylu}). Furthermore,
fluctuations of gauge field are still presented
in~(\ref{atux}). Therefore the Maxwell equation
$\nabla_{a}(X^{abcd}F_{cd})=0$ reads
\begin{equation}
\label{4dat}
\partial_{u}\left(\frac{e^{\chi(u)/2}}{u^{2}}X_{3}A^{\prime}_{t}
\right) -\frac{e^{\chi(u)/2}X_{1}}{R^{2}_{0}g(u)}(\omega
qA_{x}+q^{2}A_{t})=0,
\end{equation}
\begin{equation}
A^{\prime}_{t}+\frac{e^{-\chi(u)}g(u)u^{2}}{R^{2}_{0}}\frac{qX_{5}}{\omega
X_{3}}A^{\prime}_{x}=0,
\end{equation}
\begin{equation}
\partial_{u}(e^{-\chi(u)/2}g(u)X_{5}A^{\prime}_{x})+
\frac{e^{\chi(u)/2}}{g(u)}
X_{1}(\omega^{2}A_{x}+\omega qA_{t})=0,
\end{equation}
\begin{equation}
\partial_{u}(e^{-\chi(u)/2}g(u)X_{6}A^{\prime}_{y})+
\frac{e^{\chi(u)/2}}{g(u)}X_{2}\omega^{2}A_{y}
-\frac{e^{-\chi(u)/2}u^{2}}{R^{2}_{0}}X_{4}q^{2}A_{y}=0.
\end{equation}
The equations of motion for $B_{a}$ can be simply obtained by
replacing $A_{a}\rightarrow B_{a}$ and
$X_{i}\rightarrow\hat{X}_{i}$. In addition, the components of the
$\varepsilon$ tensor are listed below
\begin{eqnarray}
&&{\varepsilon_{tx}}^{yu}=e^{-\chi(u)/2}g(u),~~~{\varepsilon_{tu}}^
{xy}=-e^{-\chi(u)/2}g(u),\nonumber\\
& &{\varepsilon_{tu}}^{xy}=e^{-\chi(u)/2}\frac{u^{2}}{R^{2}_{0}},~~~
{\varepsilon_{xy}}^{tu}=-e^{\chi(u)/2}\frac{R_{0}^{2}}{u^{2}},
\nonumber\\
& &{\varepsilon_{xu}}^{ty}=
\frac{e^{\chi(u)/2}}{g(u)},~~~{\varepsilon_{yu}}^{tx}
=-\frac{e^{\chi(u)/2}}{g(u)}.
\end{eqnarray}
Then we can explicitly work out the relation between
$F_{ab}$ and $G_{ab}$,
\begin{eqnarray}
& &F_{tx}=g^{2}_{4}\frac{e^{-\chi(u)/2}}{X_{1}}
g(u)G_{yu},~~~F_{ty}=-g^{2}_{4}
\frac{e^{-\chi(u)/2}}{X_{2}}g(u)G_{xu},\nonumber\\
& &F_{tu}=g^{2}_{4}
\frac{e^{-\chi(u)/2}u^{2}}{R^{2}_{0}X_{3}}G_{xy},~~~
F_{xy}=-g^{2}_{4}\frac{R^{2}_{0}e^{-\chi(u)/2}}{u^{2}
X_{4}}G_{tu},\nonumber\\
& &F_{xu}=g^{2}_{4}\frac{e^{\chi(u)/2}}{g(u)X_{5}}G_{ty},~~~
F_{yu}=-g^{2}_{4}\frac{e^{\chi(u)/2}}{g(u)X_{6}}G_{tx}.
\end{eqnarray}

The boundary action can be written as follows
\begin{equation}
I_{b}=\frac{1}{2g^{2}_{4}}\int
d^{4}x\left(e^{\chi(u)/2}\frac{R^{2}_{0}}{u^{2}}
X_{3}A_{t}A^{\prime}_{t}-e^{-\chi(u)/2}g(u)X_{5}A_{x}A^{\prime}_{x}-
e^{-\chi(u)/2}g(u)X_{6}A_{y}A^{\prime}_{y}\right)|_{u\rightarrow
u_{b}}.
\end{equation}
Thus the retarded Green's functions are given by~\cite{Son:2002sd}
\begin{equation}
G^{R}_{tt}=\frac{R^{2}_{0}}{g^{2}_{4}}
\frac{e^{\chi(u)/2}X_{3}}{u^{2}}
\frac{\delta A^{\prime}_{t}}{\delta
A^{b}_{t}}|_{u\rightarrow u_{b}},
\end{equation}
\begin{equation}
G^{R}_{xx}=-\frac{1}{g^{2}_{4}}e^{-\chi(u)/2}g(u)X_{5}\frac{\delta
A^{\prime}_{x}} {\delta A^{b}_{x}}|_{u\rightarrow u_{b}},
\end{equation}
\begin{equation}
G^{R}_{tx}=\frac{1}{2g^{2}_{4}}
[\frac{e^{\chi(u)}/2}{u^{2}}X_{3}\frac{\delta
A^{\prime}_{t}}{\delta A^{b}_{x}} -e^{-\chi(u)/2}g(u)X_{5}
\frac{\delta
A^{\prime}_{x}}{\delta A^{b}_{x}}]|_{u\rightarrow u_{b}},
\end{equation}
\begin{equation}
G^{R}_{yy}=-\frac{1}{g^{2}_{4}}e^{-\chi(u)/2}g(u)X_{6}\frac{\delta
A^{\prime}_{y}}{\delta A^{b}_{y}}|_{u\rightarrow u_{b}}.
\end{equation}

Let us focus on the $yy$-component of the retarded Green's
function. The solution for $A_{y}(u)$ can be written in an abstract
form $A_{y}(u)=\psi(u)A^{b}_{y}$, where $A^{b}_{y}$ denotes its
boundary value. Therefore it can be easily seen that $\psi(u_{b})=1$
and
\begin{equation}
G^{R}_{yy}=-\frac{1}{g^{2}_{4}}e^{-\chi(u_{b})/2}
g(u_{b})X_{6}(u_{b})\psi^{\prime}(u_{b}).
\end{equation}
Recall that
\begin{equation}
F_{xy}=-\frac{g^{2}_{4}}{X_{4}}e^{\chi(u)/2}
\frac{R^{2}_{0}}{u^{2}}G_{tu}\,,
\end{equation}
therefore
\begin{equation}
B^{\prime}_{t}=C_{1}u^{2}e^{-\chi(u)/2}X_{4}\psi(u),
\end{equation}
where $C_{1}$ is some undetermined constant.
Moreover, the equation for $B_{t}$
can be deduced from~(\ref{4dat}),
\begin{equation}
\partial_{u}\left(\frac{e^{\chi(u)/2}}{u^{2}}
\hat{X}_{3}B^{\prime}_{t} \right)
-\frac{e^{\chi(u)/2}
\hat{X}_{1}}{R^{2}_{0}g(u)}(\omega qB_{x}+q^{2}B_{t})=0,
\end{equation}
which leads to
\begin{equation}
C_{1}=\frac{e^{\chi(u_{b})/2}
(\omega qB^{b}_{x}+q^{2}B^{b}_{t})}{R^{2}_{0}g(u_{b})X_{6}(u_{b})
\psi^{\prime}(u_{b})},
\end{equation}
where we have used the fact that $\hat{X}_{3}=1/X_{4}$ and
$\hat{X}_{1}=1/X_{6}$.
Then the retarded Green's function for $B_{t}$ is given by
\begin{equation}
\hat{G}^{R}_{tt}=\frac{R^{2}_{0}}{\hat{g}^{2}_{4}}
\frac{e^{\chi(u)/2}\hat{X}_{3}}{u^{2}}
\frac{\delta B^{\prime}_{t}}{\delta B^{b}_{t}}=
\frac{g^{2}_{4}e^{\chi(u_{b})/2}q^{2}}
{g(u_{b})X_{6}(u_{b})\psi^{\prime}(u_{b})}.
\end{equation}
Finally we arrive at
\begin{equation}
\hat{G}^{R}_{tt}G^{R}_{yy}=-q^{2},~~
\Rightarrow~~K^{T}(\omega,q)\hat{K}^{L}(\omega,q)=1,
\end{equation}
while we can also obtain $K^{L}(\omega,q)\hat{K}^{T}(\omega,q)=1$ in a
parallel way. Our results indicate thus that such a simple duality
relation still holds in our domain-wall geometry, irrespective of the
IR near horizon geometry.

%%%%%%%%%%%%%%%%%%%%%%
\subsection{DC conductivity at zero temperature}
\label{sec3.4}
%%%%%%%%%%%%%%%%%%%%%

Up to now we have discussed the conductivity at finite temperature,
while the conductivity at extremality can be studied in a somewhat
different way. In this case the asymptotic geometry is still AdS, but
the near horizon geometry is Lifshitz metric. Notice that the Weyl
tensor vanishes in AdS spacetime, so the asymptotic solution of the
gauge field is still given by
\begin{equation}
A_{y}=A^{(0)}_{y}+\frac{A^{(1)}_{y}}{r^{d-1}}\,.
\end{equation}
It was observed in~\cite{Horowitz:2009ij} that the equation of motion
for $A_{y}$ can be recast into a Schr\"{o}dinger equation
\begin{equation}
-A_{y,ss}+V(s)A_{y}=\omega^{2}A_{y}\,,
\end{equation}
where $s$ denotes some
redefinition of the radial coordinate. The conductivity can be
expressed in terms of the reflection coefficient $\mathcal{R}$
\begin{equation}
\sigma=\frac{1-\mathcal{R}}{1+\mathcal{R}}.
\end{equation}

The general strategy can be summarized as follows: we solve the
Schr\"{o}dinger equation in the near horizon region and the asymptotic
region respectively and then match the two solutions in certain
intermediate region. Thus the reflection coefficient can be determined
and the conductivity is obtained. In our specific background, let us
consider the four-dimensional case as an example. Recall that the
equation of motion for $A_{y}$ is given by
\begin{equation*}
\partial_{r}[e^{-\chi(r)/2}g(r)G(r)A^{\prime}_{y}]+
\frac{e^{\chi(r)/2}}{g(r)}G(r)\omega^{2}A_{y}=0,
~~~G(r)=1-\frac{\gamma L^{2}}{3r^{2}}F(r).
\end{equation*}
By introducing
\begin{equation}
\frac{\partial}{\partial s}=e^{-\chi/2}g
\frac{\partial}{\partial r},~~~
\Psi=\sqrt{G(r)}A_{y},
\end{equation}
the above equation turns out to be of Schr\"{o}dinger form
\begin{equation}
-\partial_{s}^{2}\Psi+V(s)\Psi=\omega^{2}\Psi,~~~V(s)
=\frac{1}{\sqrt{G(r)}}\partial^{2}_{s}\sqrt{G(r)}.
\end{equation}
However, it can be seen that
\begin{equation}
G(r)_{\rm IR}=1-\frac{1}{3}\gamma L^{2},~~~G(r)_{\rm UV}=1,
\end{equation}
which leads to a trivial potential $V(s)=0$.
Therefore we can easily obtain $\mathcal{R}=0$ and $\sigma=1$.

%%%%%%%%%%%%%%%%%%%%%%%%%%%%%%%%%%%%%%%
\section{Charge transport in higher dimensions}
\label{sec4}
%%%%%%%%%%%%%%%%%%%%%%%%%%%%%%%%%%%%%%%%
In this section we calculate the conductivity in a general
$(d+2)$-dimensional spacetime, where we apply the same techniques
adopted in section 3. It was observed
in~\cite{Kovtun:2008kx} that in general $(d+2)$-dimensional
background, the electrical conductivity and charge susceptibility are
fixed by the central charge in a universal manner. However, due to our
lack of understanding on the conformal field theory side, the
relations between the conductivity and the central charge are still
unclear. Furthermore, conductivity in asymptotically Lifshitz
spacetimes was also studied in~\cite{Pang:2009wa}, where the focus
was on the leading order effective action. In addition, since
higher-dimensional electro-magnetic duality is not so powerful
as its four-dimensional counterparts, we will not consider it.
%%%%%%%%%%%%%%%%%%%%%%%%%%%%%%%%%%%%%%%%%%%%
\subsection{DC Conductivity from the membrane paradigm}
\label{sec4.1}
%%%%%%%%%%%%%%%%%%%%%%%%%%%%%%%%%%%%%%%%%%%%%
Considering the following $(d+2)$-dimensional domain-wall geometry
\begin{equation}
\label{d2all}
ds^{2}=-g(r)e^{-\chi(r)}dt^{2}+\frac{dr^{2}}{g(r)}+
\frac{r^{2}}{R_{0}^2}\sum\limits^{d}_{i=1}dx^{2}_{i},
\end{equation}
whose IR near horizon metric is given by
\begin{equation}
\label{d2IR}
ds^{2}_{\rm
IR}=-\frac{r^{2z}}{L^{2}}f(r)dt^{2}+\frac{L^{2}dr^{2}}{r^{2}f(r)}
+\frac{r^{2}}{L^{2}}\sum\limits^{d}_{i=1}dx^{2}_{i},
~~~f(r)=1-\frac{r_{0}^{z+d}}{r^{z+d}}\,.
\end{equation}
It becomes Schwarzschild-${\rm AdS}_{d+2}$ when $z=1$. It can be
seen that here we still have
\begin{equation}
e^{-\chi(r)}=r^{2z-2},~~~g(r)=\frac{r^{2}f(r)}{L^{2}},~~~R_{0}=L,
\end{equation}
The UV geometry is still fixed to be AdS.  In the
background~(\ref{d2all}), the non-vanishing components of the Weyl
tensor are given as follows
\begin{eqnarray}
& &C_{trtr}=\frac{(d-1)e^{-\chi(r)}}{4(d+1)r^{2}}F(r),~~~
C_{titj}=-\frac{(d-1)e^{-\chi(r)}}{4d(d+1)R_{0}^2}g(r)
F(r)\delta_{ij},\nonumber\\
& &C_{rirj}=\frac{d-1}{4d(d+1)R_{0}^{2}}\frac{F(r)}{g(r)}
\delta_{ij},~~~
C_{ijkl}=-\frac{r^{2}}{2d(d+1)R_{0}^{4}}F(r)\delta_{ik}
\delta_{jl},
\end{eqnarray}
where $i,j,k,l=x_{1},\cdots,x_{d}$ and $F(r)$ is still given
by~(\ref{fr}).  Following the procedures exhibited in section
3, we obtain the conductivity
\begin{eqnarray}
\sigma&=&\frac{1}{g^{2}_{d+2}}
\sqrt{-g}\sqrt{-X^{txtx}X^{rxrx}}|_{r=r_{0}},\nonumber\\
&=&\frac{1}{g^{2}_{d+2}}\left(\frac{r_{0}}{L}\right)^{d-2}
\left[1-\frac{4(d-1)\gamma}{d(d+1)}
(2z(z-1)+d(z-d-2))\right]\,.
\end{eqnarray}
This reduces to~(\ref{dc4}) when $d=2$.
%%%%%%%%%%%%%%%%%%%%%%%%%%%%%%%%%%%%%%%%%
\subsection{DC Conductivity from Kubo's formula}
\label{sec4.2}
%%%%%%%%%%%%%%%%%%%%%%%%%%%%%%%%%%%%%%%%%%
To evaluate the conductivity from Kubo's formula,
we introduce a new radial coordinate $u$,
\begin{equation}
\label{duzb}
ds^{2}=-g(u)e^{-\chi(u)}dt^{2}+\frac{du^{2}}{g(u)}+
\frac{R_{0}^{2}}{u^{2}}\sum\limits^{d}_{i=1}dx^{2}_{i}.
\end{equation}
The IR metric can be written as follows in the $u$-coordinate
\begin{equation}
ds^{2}_{\rm
IR}=-\frac{r_{0}^{2z}}{L^{2}u^{2z}}f(u)dt^{2}+\frac{L^{2}du^{2}}{u^{2}f(u)}
+\frac{r_{0}^{2}}{L^{2}u^{2}}\sum\limits^{d}_{i=1}dx^{2}_{i},~~~f(u)=1-u^{z+d},
\end{equation}
where the horizon locates at $u=1$. Comparing the two metrics we can
obtain
\begin{equation}
e^{-\chi(u)}=\frac{r_{0}^{2z}}{u^{2z+2}},~~~g(u)=
\frac{u^{2}f(u)}{L^{2}},~~~R_{0}=\frac{r_{0}}{L},
\end{equation}
The corresponding non-vanishing components of the Weyl tensor are given by
\begin{eqnarray}
\label{weylud}
& &C_{tutu}=\frac{(d-1)e^{-\chi(u)}}{4(d+1)u^{2}}F(u),~~~
C_{titj}=-\frac{(d-1)R_{0}^{2}e^{-\chi(u)}}{4d(d+1)
u^{4}}g(u)F(u)\delta_{ij},\nonumber\\
& &C_{uiuj}=\frac{(d-1)R_{0}^{2}}{4d(d+1)u^{4}}
\frac{F(u)}{g(u)}\delta_{ij},~~~
C_{ijkl}=-\frac{R_{0}^{4}}{2d(d+1)u^{6}}F(u)
\delta_{ik}\delta_{jl},
\end{eqnarray}
where $i,j,k,l=x,y$ and $F(u)$ is still given
by~(\ref{fu}). Hence the generalized Maxwell equation
in $(d+2)$-dimensions reads
\begin{equation}
\label{ayd2}
A^{\prime\prime}_{y}+
\frac{M^{\prime}_{d+2}(u)}{M_{d+2}(u)}A^{\prime}_{y}+
\frac{e^{\chi(u)}}{g(u)^{2}}
\omega^{2}A_{y}=0,
\end{equation}
where
\begin{equation}
M_{d+2}(u)=\left(1-\frac{2
\gamma (d-1)L^{2}}{d(d+1)u^{2}}F(u)\right)
\frac{e^{-\chi(u)/2}}{u^{d-2}}g(u).
\end{equation}
On the other hand, the retarded Green's function turns out to be
\begin{equation}
G^{R}_{yy}=-\frac{1}{g^{2}_{4}}\sqrt{-g}
g^{uu}g^{yy}(1-8\gamma L^{2}{C_{uy}}^{uy})
\frac{A_{y}(u,-q)\partial_{u}
A_{y}(u,q)}{A_{y}(u,-q)A_{y}(u,q)}\big|_{u\rightarrow u_{b}}.
\end{equation}
Therefore one can find that
\begin{equation}
\sqrt{-g}g^{uu}g^{yy}(1-8
\gamma L^{2}{C_{uy}}^{uy})=R^{d-2}_{0}M_{d+2}(u),
\end{equation}

For our general $(d+2)$-dimensional case, the conserved quantity
in~(\ref{qu}) is given by
\begin{equation}
Q(u)=M_{d+2}(u)(\bar{A}_{y}\partial_{u}A_{y}-
A_{y}\partial_{u}\bar{A}_{y}),
\end{equation}
which leads to the following expression for
the retarded Green's function
\begin{equation}
{\rm Im}G^{R}_{yy}=-\frac{R_{0}^{d-2}}{2ig^{2}_{d+2}}
\frac{Q(u)}{|A_{y}(u_{b})|^{2}}.
\end{equation}
Furthermore, the general solution to $A_{y}$ can still be written as
\begin{equation}
A_{y}(u)=(1-u)^{-\frac{i\omega}{4\pi T}}y(u).
\end{equation}
Thus we can obtain
\begin{equation}
{\rm Im}G^{R}_{yy}=-\frac{\omega
R^{d-2}_{0}}{g^{2}_{d+2}}\left[1-\frac{4(d-1)\gamma}{d(d+1)}
(2z(z-1)+d(z-d-2))\right]\frac{|y(1)|^{2}}{|y(u_{b})|^{2}}.
\end{equation}
Finally, in the low frequency limit the solution to~(\ref{ayd2}) is
simply $y(u)=\rm const$, which results in
\begin{equation}
\sigma=-\frac{1}{\omega}{\rm Im}G^{R}_{yy}=
\frac{r_{0}^{d-2}}{g^{2}_{d+2}L^{d-2}}
\left[1-\frac{4(d-1)\gamma}{d(d+1)}(2z(z-1)+d(z-d-2))\right].
\end{equation}
It can be seen that once again this result agrees
with the one obtained via the membrane paradigm.
%%%%%%%%%%%%%%%%%%%%%%%%%%%%%%%%%%%%%%%%%%%%%%%%%%%%%%%%%%%%
\section{Summary and discussion}
\label{sec5}
%%%%%%%%%%%%%%%%%%%%%%%%%%%%%%%%%%%%%%%%%%%%%%%%%%%%%%%%%%%%
The full background geometry is required when
calculating the retarded Green's functions via AdS/CFT. However, we
can still acquire some knowledge about the transport coefficients from
a domain-wall geometry. In this paper we computed conductivity
in the presence of Weyl corrections in a domain-wall background, whose
near horizon IR geometry is Lifshitz black hole and asymptotic
geometry is AdS. We obtained the conductivity via both the membrane
paradigm and Kubo's formula. By making use of a shortcut, the
conductivity derived from Kubo's formula can be solely expressed in
terms of quantities at the horizon. The results obtained via both
approaches precisely match in four as well as in higher
dimensions. Moreover, it was shown in~\cite{Myers:2010pk} that in
four dimensions, although self-duality was lost in higher derivative
theories, a simple relation for the longitudinal and transverse
components of the current-current correlation functions and those of
the dual counterparts, $K^{L}(\omega,q)\hat{K}^{T}(\omega,q)=1$, still
held. Here we show that this simple relation also holds in our
domain-wall background, irrespective of the IR near horizon geometry.

Similar backgrounds were also investigated in~\cite{Goldstein:2009cv}
and~\cite{Chen:2010kn}, where the authors considered charged dilaton
black branes in Einstein-Maxwell-Dilaton theory, whose near horizon
geometry was Lifshitz metric and asymptotic geometry was AdS. One
crucial difference was that due to the nontrivial background $U(1)$
gauge field, the potential in the Schr\"{o}dinger equation was also
nontrivial, which lead to a universal conductivity ${\rm
Re}\sigma\sim\omega^{2}$ in four dimensions. If we want to consider
Weyl corrections to the conductivity in such a background, it would be
necessary to work out the perturbed metric, as the nontrivial
background gauge field would back-react on the leading order
solution. Holographic properties of charged black holes in higher
derivative theories were studied in~\cite{Myers:2009ij,
Cremonini:2008tw, Cai:2011uh} and transport properties in extremal
charged black hole backgrounds were considered in~\cite{Paulos:2009yk,
extremal}.

One can also consider the following type of higher order corrections
instead
\begin{equation}
\label{equ4}
I^{\prime}_{\rm vec}=\frac{1}{\tilde{g}^{2}_{4}}
\int d^{4}x\sqrt{-g}[-\frac{1}{4}F_{ab}F^{ab}+
\alpha L^{2}(R_{abcd}F^{ab}F^{cd}-
4R_{ab}F^{ac}{F^{b}}_{c}+RF^{ab}F_{ab})],
\end{equation}
which arises from the Kaluza-Klein reduction of five-dimensional
Gauss-Bonnet gravity.  It was observed in~\cite{Myers:2010pk} that by
combining the Einstein equation in the neutral black hole background
$R_{ab}=-3/L^{2}g_{ab}$ and the definition of the Weyl tensor, the
action~(\ref{equ4}) becomes
\begin{equation}
I^{\prime}_{\rm vec}=\frac{1+8\alpha}{\tilde{g}^{2}_{4}}
\int d^{4}x\sqrt{-g}[-\frac{1}{4}F_{ab}F^{ab}
+\frac{\alpha}{1+8\alpha}L^{2}C_{abcd}F^{ab}F^{cd}].
\end{equation}
It can be easily seen that the resulting action is
equivalent to~(\ref{equ2}) with the following identifications
\begin{equation}
g^{2}_{4}=\frac{\tilde{g}^{2}_{4}}{1+8
\alpha},~~~\gamma=\frac{\alpha}{1+8\alpha}.
\end{equation}
Therefore the charge transport properties are identical. However, here
the Einstein equation in the IR Lifshitz black hole background cannot
have such a simple expression and thus the two actions are generically
not equivalent. It would be interesting to study charge transport
coefficients in a different theory e.g.~(\ref{equ4}) and to see the
effects of higher order corrections on the conductivity.

\bigskip \goodbreak \centerline{\bf Acknowledgments}
\noindent  DWP acknowledges an FCT
(Portuguese Science Foundation) grant. This work was also funded by
FCT through projects CERN/FP/109276/2009 and PTDC/FIS/098962/2008.

%%%%%%%%%%%%%%%%%%%%%%%%%%%%%%%%%%%%%%%%%%%%%%%%%
%\appendix
%%%%%%%%%%%%%%%%%%%%%%%%%%%%%%%%%%%%%%%%%%%%%%%%%%
%\section{Black D2 in dual frame}

%%%%%%%%%%%%%%%%%%%%%%%%%%%%%%%%%%%%%%%%%%%%%%%%%%


\begin{thebibliography}{99}
%%%%%%%%%%%%%%%%%%%%%%%%%%%%%%%%%%%%%%%%%%%%%%%%%%
\addcontentsline{toc}{section}{References}
%%%%%%%%%%%%%%%%%%%%%%%%%%%%%%%%%%%%%%%%%%%%%%%%%%


\bibitem{Maldacena:1997re}
   J.~M.~Maldacena,
   ``The large N limit of superconformal field theories
   and supergravity'',
   Adv.\ Theor.\ Math.\ Phys.\  {\bf 2}, 231 (1998)
   [Int.\ J.\ Theor.\ Phys.\  {\bf 38}, 1113 (1999)]
   [arXiv:hep-th/9711200].\\
   S.~S.~Gubser, I.~R.~Klebanov and A.~M.~Polyakov,
   ``Gauge theory correlators from non-critical string theory'',
   Phys.\ Lett.\  B {\bf 428}, 105 (1998)
   [arXiv:hep-th/9802109].\\
   E.~Witten,
   ``Anti-de Sitter space and holography'',
   Adv.\ Theor.\ Math.\ Phys.\  {\bf 2}, 253 (1998)
   [arXiv:hep-th/9802150].

\bibitem{Aharony:1999ti}
   O.~Aharony, S.~S.~Gubser, J.~M.~Maldacena, H.~Ooguri and Y.~Oz,
   ``Large N field theories, string theory and gravity'',
   Phys.\ Rept.\  {\bf 323}, 183 (2000)
   [arXiv:hep-th/9905111].

\bibitem{Hartnoll:2009sz}
   S.~A.~Hartnoll,
   ``Lectures on holographic methods for condensed matter physics'',
   Class.\ Quant.\ Grav.\  {\bf 26}, 224002 (2009)
   [arXiv:0903.3246 [hep-th]].\\
   C.~P.~Herzog,
   ``Lectures on Holographic Superfluidity and Superconductivity'',
   J.\ Phys.\ A  {\bf 42}, 343001 (2009)
   [arXiv:0904.1975 [hep-th]].\\
    J.~McGreevy,
   ``Holographic duality with a view toward many-body physics'',
   arXiv:0909.0518 [hep-th].\\
   G.~T.~Horowitz,
   ``Introduction to Holographic Superconductors'',
   arXiv:1002.1722 [hep-th].\\
   S.~Sachdev,
   ``Condensed matter and AdS/CFT'',
   arXiv:1002.2947 [hep-th].
\bibitem{Son:2008ye}
   D.~T.~Son,
   ``Toward an AdS/cold atoms correspondence:
    A Geometric realization of the
   Schrodinger symmetry'',
   Phys.\ Rev.\  D {\bf 78}, 046003 (2008)
   [arXiv:0804.3972 [hep-th]].
\bibitem{Balasubramanian:2008dm}
   K.~Balasubramanian and J.~McGreevy,
   ``Gravity duals for non-relativistic CFTs'',
   Phys.\ Rev.\ Lett.\  {\bf 101}, 061601 (2008)
   [arXiv:0804.4053 [hep-th]].
\bibitem{Kachru:2008yh}
   S.~Kachru, X.~Liu and M.~Mulligan,
   ``Gravity Duals of Lifshitz-like Fixed Points'',
   Phys.\ Rev.\  D {\bf 78}, 106005 (2008)
   [arXiv:0808.1725 [hep-th]].
\bibitem{Herzog:2007ij}
   C.~P.~Herzog, P.~Kovtun, S.~Sachdev and D.~T.~Son,
   ``Quantum critical transport, duality, and M-theory'',
   Phys.\ Rev.\  D {\bf 75}, 085020 (2007)
   [arXiv:hep-th/0701036].

\bibitem{Myers:2010pk}
   R.~C.~Myers, S.~Sachdev and A.~Singh,
   ``Holographic Quantum Critical Transport without Self-Duality'',
   arXiv:1010.0443 [hep-th].

\bibitem{Lemos:1994fn}
   J.~P.~S.~Lemos, ``Two-dimensional black holes and planar general
   relativity'', Class. Quant. Grav. {\bf 12}, 1081 (1995)
   [arXiv:gr-qc/9407024].\\
   J.~P.~S.~Lemos, ``Cylindrical black hole in general relativity'',
   Phys. Lett. B {\bf 353}, 46 (1995) [arXiv:gr-qc/9404041].\\
   J.~P.~S.~Lemos and V.~T.~Zanchin, ``Rotating Charged Black String
   and Three Dimensional Black Holes'', Phys.\ Rev.\ D {\bf 54}, 3840
   (1996) [arXiv:hep-th/9511188].

\bibitem{Taylor:2008tg}
   M.~Taylor,
   ``Non-relativistic holography'',
   arXiv:0812.0530 [hep-th].
\bibitem{Lif}
   U.~H.~Danielsson and L.~Thorlacius,
   ``Black holes in asymptotically Lifshitz spacetime'',
   JHEP {\bf 0903}, 070 (2009)
   [arXiv:0812.5088 [hep-th]].\\
   R.~B.~Mann,
   ``Lifshitz Topological Black Holes'',
   JHEP {\bf 0906}, 075 (2009)
   [arXiv:0905.1136 [hep-th]].\\
   D.~W.~Pang,
   ``A Note on Black Holes in Asymptotically Lifshitz Spacetime'',
   [arXiv:0905.2678 [hep-th]].\\
   G.~Bertoldi, B.~A.~Burrington and A.~Peet,
   ``Black Holes in asymptotically Lifshitz spacetimes
    with arbitrary critical
   exponent'',
   Phys.\ Rev.\  D {\bf 80}, 126003 (2009)
   [arXiv:0905.3183 [hep-th]].\\
   G.~Bertoldi, B.~A.~Burrington and A.~W.~Peet,
   ``Thermodynamics of black branes in asymptotically
   Lifshitz spacetimes'',
   Phys.\ Rev.\  D {\bf 80}, 126004 (2009)
   [arXiv:0907.4755 [hep-th]].\\
   K.~Balasubramanian and J.~McGreevy,
   ``An Analytic Lifshitz black hole'',
   Phys.\ Rev.\  D {\bf 80}, 104039 (2009)
   [arXiv:0909.0263 [hep-th]].\\
   E.~Ayon-Beato, A.~Garbarz, G.~Giribet and M.~Hassaine,
   ``Lifshitz Black Hole in Three Dimensions'',
   Phys.\ Rev.\  D {\bf 80}, 104029 (2009)
   [arXiv:0909.1347 [hep-th]].\\
   R.~G.~Cai, Y.~Liu and Y.~W.~Sun,
   ``A Lifshitz Black Hole in Four Dimensional $R^2$ Gravity'',
   JHEP {\bf 0910}, 080 (2009)
   [arXiv:0909.2807 [hep-th]].\\
   E.~Ayon-Beato, A.~Garbarz, G.~Giribet and M.~Hassaine,
   ``Analytic Lifshitz black holes in higher dimensions'',
   JHEP {\bf 1004}, 030 (2010)
   [arXiv:1001.2361 [hep-th]].\\
   G.~Bertoldi, B.~A.~Burrington and A.~W.~Peet,
   ``Thermal behavior of charged dilatonic black branes in AdS and UV
   completions of Lifshitz-like geometries'',
   Phys.\ Rev.\  D {\bf 82}, 106013 (2010)
   [arXiv:1007.1464 [hep-th]].
\bibitem{Brigante:2007nu}
   M.~Brigante, H.~Liu, R.~C.~Myers, S.~Shenker and S.~Yaida,
   ``Viscosity Bound Violation in Higher Derivative Gravity'',
   Phys.\ Rev.\  D {\bf 77}, 126006 (2008)
   [arXiv:0712.0805 [hep-th]].
\bibitem{Ritz:2008kh}
   A.~Ritz, J.~Ward,
   ``Weyl corrections to holographic conductivity'',
   Phys.\ Rev.\  {\bf D79}, 066003 (2009).
   [arXiv:0811.4195 [hep-th]].
\bibitem{Kovtun:2003wp}
   P.~Kovtun, D.~T.~Son and A.~O.~Starinets,
   ``Holography and hydrodynamics: Diffusion on stretched horizons'',
   JHEP {\bf 0310}, 064 (2003)
   [arXiv:hep-th/0309213].
\bibitem{Iqbal:2008by}
   N.~Iqbal and H.~Liu,
   ``Universality of the hydrodynamic limit in AdS/CFT and the membrane
   paradigm'',
   Phys.\ Rev.\  D {\bf 79}, 025023 (2009)
   [arXiv:0809.3808 [hep-th]].
\bibitem{Son:2002sd}
   D.~T.~Son, A.~O.~Starinets,
   ``Minkowski space correlators in AdS / CFT correspondence:
    Recipe and applications'',
   JHEP {\bf 0209}, 042 (2002).
   [arXiv:hep-th/0205051].
\bibitem{CaronHuot:2006te}
   S.~Caron-Huot, P.~Kovtun, G.~D.~Moore, A.~Starinets and L.~G.~Yaffe,
   ``Photon and dilepton production in supersymmetric Yang-Mills plasma'',
   JHEP {\bf 0612}, 015 (2006)
   [arXiv:hep-th/0607237].
\bibitem{Atmaja:2008mt}
   A.~Nata Atmaja and K.~Schalm,
   ``Photon and Dilepton Production in Soft Wall AdS/QCD'',
   JHEP {\bf 1008}, 124 (2010)
   [arXiv:0802.1460 [hep-th]].
\bibitem{Kovtun:2008kx}
   P.~Kovtun and A.~Ritz,
   ``Universal conductivity and central charges'',
   Phys.\ Rev.\  D {\bf 78}, 066009 (2008)
   [arXiv:0806.0110 [hep-th]].
\bibitem{Pang:2009wa}
   D.~W.~Pang,
   ``Conductivity and Diffusion Constant in Lifshitz Backgrounds'',
   JHEP {\bf 1001}, 120 (2010)
   [arXiv:0912.2403 [hep-th]].
\bibitem{Horowitz:2009ij}
   G.~T.~Horowitz and M.~M.~Roberts,
   ``Zero Temperature Limit of Holographic Superconductors'',
   JHEP {\bf 0911}, 015 (2009)
   [arXiv:0908.3677 [hep-th]].
\bibitem{Goldstein:2009cv}
   K.~Goldstein, S.~Kachru, S.~Prakash and S.~P.~Trivedi,
   ``Holography of Charged Dilaton Black Holes'',
   JHEP {\bf 1008}, 078 (2010)
   [arXiv:0911.3586 [hep-th]].
\bibitem{Chen:2010kn}
   C.~M.~Chen and D.~W.~Pang,
   ``Holography of Charged Dilaton Black Holes in General Dimensions'',
   JHEP {\bf 1006}, 093 (2010)
   [arXiv:1003.5064 [hep-th]].
\bibitem{Myers:2009ij}
   R.~C.~Myers, M.~F.~Paulos and A.~Sinha,
   ``Holographic Hydrodynamics with a Chemical Potential'',
   JHEP {\bf 0906}, 006 (2009)
   [arXiv:0903.2834 [hep-th]].
\bibitem{Cremonini:2008tw}
   S.~Cremonini, K.~Hanaki, J.~T.~Liu and P.~Szepietowski,
   ``Black holes in five-dimensional gauged supergravity with higher
   derivatives'',
   JHEP {\bf 0912}, 045 (2009)
   [arXiv:0812.3572 [hep-th]].
\bibitem{Cai:2011uh}
   R.~G.~Cai and D.~W.~Pang,
   ``Holography of Charged Black Holes with $RF^2$ Corrections'',
   arXiv:1104.4453 [hep-th].
\bibitem{Paulos:2009yk}
   M.~F.~Paulos,
   ``Transport coefficients, membrane couplings
     and universality at extremality'',
   JHEP {\bf 1002}, 067 (2010).
   [arXiv:0910.4602 [hep-th]].
\bibitem{extremal}
   M.~Edalati, J.~I.~Jottar, R.~G.~Leigh,
   ``Transport Coefficients at Zero Temperature
     from Extremal Black Holes'',
   JHEP {\bf 1001}, 018 (2010).
   [arXiv:0910.0645 [hep-th]].\\
   R.~G.~Cai, Y.~Liu, Y.~W.~Sun,
   ``Transport Coefficients from Extremal Gauss-Bonnet Black Holes'',
   JHEP {\bf 1004}, 090 (2010).
   [arXiv:0910.4705 [hep-th]].\\
   S.~K.~Chakrabarti, S.~Jain, S.~Mukherji,
   ``Viscosity to entropy ratio at extremality'',
   JHEP {\bf 1001}, 068 (2010).
   [arXiv:0910.5132 [hep-th]].\\
   M.~Edalati, J.~I.~Jottar, R.~G.~Leigh,
   ``Shear Modes, Criticality and Extremal Black Holes'',
   JHEP {\bf 1004}, 075 (2010).
   [arXiv:1001.0779 [hep-th]].


\end{thebibliography}
\end{document}